\documentclass[sigconf]{acmart}

\usepackage{acmart-taps}
\usepackage{xspace, balance}

\copyrightyear{2026}
\acmYear{2026}
\setcopyright{cc}
\setcctype{by}
\acmConference[UIST Adjunct '26]{The 39th Annual ACM Symposium on User Interface Software and Technology}{November 02--05, 2026}{Detroit, MI, USA}
\acmBooktitle{The 39th Annual ACM Symposium on User Interface Software and Technology (UIST Adjunct '26), November 02--05, 2026, Detroit, MI, USA}
\acmDOI{10.1145/3830397.3841841}
\acmISBN{979-8-4007-2855-6/2026/11}


\begin{document}

\author{Nishanth Chidambaram}
\orcid{0009-0008-2729-5626}
\affiliation{%
  \institution{University of California San Diego}
  \city{La Jolla}
  \state{CA}
  \country{USA}
}
\email{nchidambaram@ucsd.edu}

\author{Kaustubh Paliwal}
\orcid{0009-0000-3351-388X}
\affiliation{%
  \institution{University of California San Diego}
  \city{La Jolla}
  \state{CA}
  \country{USA}
}
\email{kpaliwal@ucsd.edu}

\author{Kayla Hom}
\orcid{0009-0007-0729-2914}
\affiliation{%
  \institution{University of California San Diego}
  \city{La Jolla}
  \state{CA}
  \country{USA}
}
\email{krhom@ucsd.edu}

\author{Shaoze Zhou}
\orcid{0009-0000-3243-0599}
\affiliation{%
  \institution{Florida International University}
  \city{Miami}
  \state{FL}
  \country{USA}
}
\email{szhou010@fiu.edu}

\author{Chen Chen}
\orcid{0000-0001-7179-0861}
\affiliation{%
  \institution{Florida International University}
  \city{Miami}
  \state{FL}
  \country{USA}
}
\email{chechen@fiu.edu}

\author{Manas Satish Bedmutha}
\orcid{0000-0003-3427-2226}
\affiliation{%
  \institution{University of California San Diego}
  \city{La Jolla}
  \state{CA}
  \country{USA}
}
\email{mbedmutha@ucsd.edu}

\author{Nadir Weibel}
\orcid{0000-0002-3457-4227}
\affiliation{%
  \institution{University of California San Diego}
  \city{La Jolla}
  \state{CA}
  \country{USA}
}
\email{weibel@ucsd.edu}

\renewcommand{\shortauthors}{Chidambaram \emph{et al.}}


\def\sysname{AffAdapt}

\title{AffAdapt: AFFect-driven ADAPTive AI Personas for Seamless Conversations}

\begin{abstract}
AI-generated personas are being increasingly used for support, training and simulations. While generative AI models possess abilities to generate affect-aware responses, their embodiment into visual personas is an active area of investigation. Naturalistic exchanges require understanding of the conversational partners' turn completions, whether the agent should respond or keep listening and rely on non-verbal cues aligned with one's emotional states. Seamless human-AI conversation in a multimodal setting requires all modalities being generated to act in coordination. We present \sysname, a seamless interaction design framework for AI-personas, which coordinates streaming speech recognition, proactive turn-management, persona-grounded response generation, a persistent emotional state, and synchronized embodied output into a single interaction loop. We demonstrate the architecture in the context of practicing sensitive, high-stakes conversations, and report an initial case study showing fluid turn management and adaptive, persona-consistent behavior, alongside open challenges in interruption handling, open-ended dialogue, and multimodal affective alignment. \sysname's interaction loop is a generalizable pattern for coordinating timing, identity, and affect in real-time AI personas - applicable to training, coaching, education, and simulation contexts wherever believable, responsive interaction matters.
\end{abstract}

\maketitle

\begin{figure}[t!]
    \centering
    \includegraphics[width=.99\columnwidth]{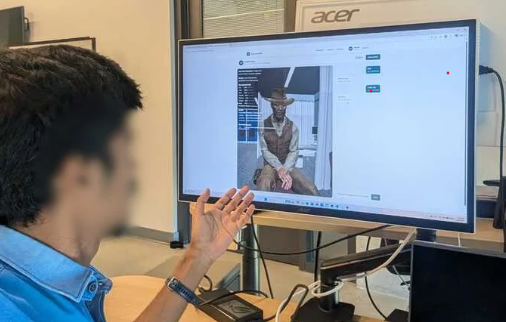}
    \caption{\sysname~in use; a user mid conversation with the agent while monitoring the emotional state of the agent.}\
    \label{fig::system}
\end{figure}

\section{Introduction}

AI personas, or conversational agents with a defined identity and communicative style, are increasingly used across training, simulation, and role-play~\cite{chen2024oscars}. This raises the interface challenge of not only generating a plausible response, but also proactively deciding when to keep listening, when a user's turn is complete, when to respond, when to yield, and how to recover from an interruption~\cite{Skantze2021}. These are timing and control decisions, not merely modeling decisions. Systems can generate strong responses and still feel unnatural if they respond at the wrong moment or cut a user off.

This problem becomes more demanding if agents must also behave using a persona. A generic voice assistant only needs to answer a query; an AI persona must do so while maintaining a consistent identity and managing its emotional state, social stance, and relationship trajectory across the conversation. Recent full-duplex speech-to-speech models and commercial real-time voice APIs~\cite{Defossez2024Moshi, OpenAIRealtimeAPI} substantially advance the underlying speech model and infrastructure for open-domain assistants, and recent HCI work has shown that supporting interruptions and backchannels improves perceived naturalness and engagement in voice agents~\cite{Liu2025Interruptions, Ding2022TalkTive}. Our focus is adjacent but distinct: the interaction-policy layer for an agent that must remain in character while making these decisions, coordinating turn management with a persona's identity and evolving emotional state, rather than advancing the underlying speech model.

We instantiate this architecture as \sysname, a prototype for practicing sensitive, high-stakes conversations. HIV prevention and care motivates our current design (trust-building conversations around disclosure and stigma are a demanding test case for persona-grounded, proactively responsive interaction), but the contribution of this work is the interaction layer that coordinates speech, persona state, dialogue generation, and embodiment, independent of any single application domain.

This poster contributes: (1) a modular interaction approach coordinating proactive turn management, persona-grounded response generation, and embodied output into a single loop; and (2) a persistent, trainee-driven persona state that conditions both dialogue and embodiment based on how the agent is treated.

\section{System Design}

\begin{figure}[t]
  \centering
  \includegraphics[width=0.6\columnwidth]{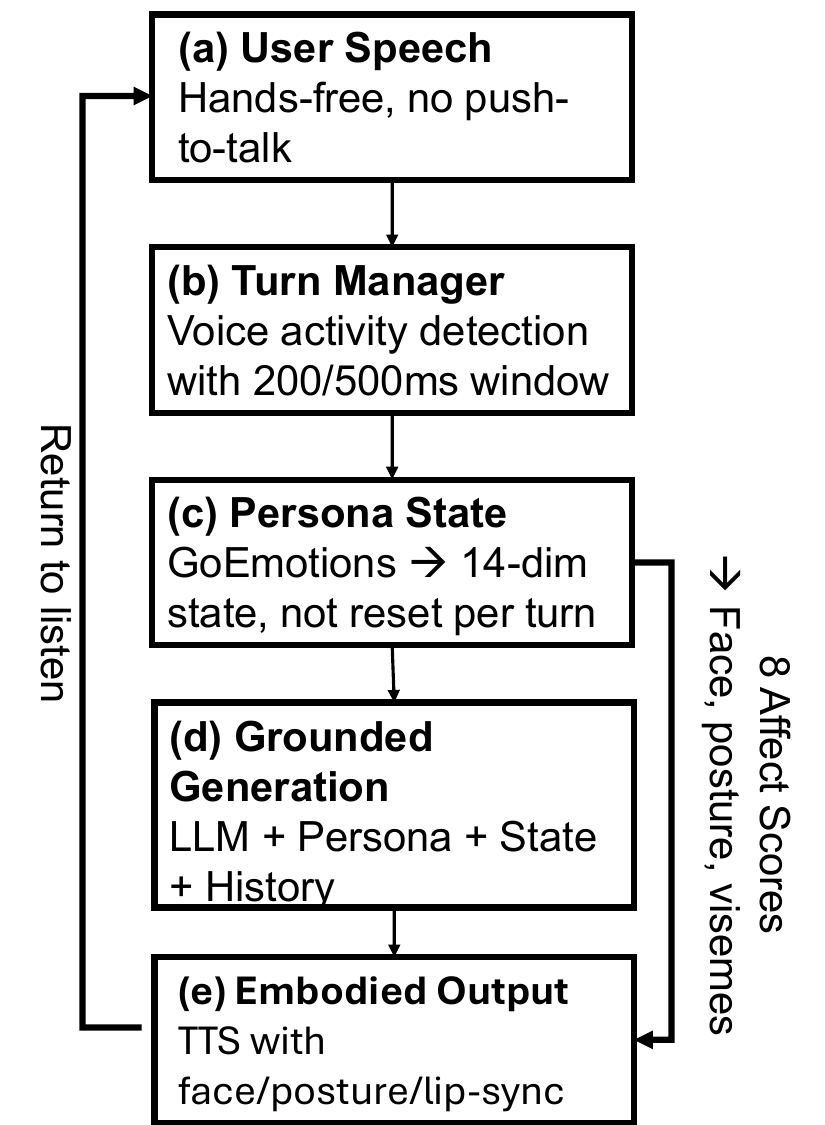}
  \caption{The \sysname\ interaction loop. The trainee's (a)~speech is segmented
  into turns by (b)~a turn manager (voice-activity detection); each completed turn
  updates (c)~a persistent fourteen-dimensional model of the persona's emotional
  state, which conditions both (d)~persona-grounded generation and (e)~the
  synchronized embodied output (voice, face, posture, lip-sync) before the loop
  returns to listening. The persona state (c)~is the hub: it drives both what the
  agent says (d) and how it appears (e), so the agent adapts to how it is treated
  rather than merely responding quickly.}
  \label{fig::system}
\end{figure}

\sysname\ treats seamlessness as a coordination problem across modules (turn
management, persona state, dialogue generation, and embodiment) rather than a
property of any single model or channel, as shown in Figure~\ref{fig::system}.

\vspace{0.25em}
\noindent
\textbf{Proactive Turn Management.}
The loop couples streaming recognition (\textit{faster-whisper}~\cite{Radford2023Whisper}, medium), a large language model, and neural text-to-speech (\textit{Kokoro}~\cite{HexgradKokoro}). Rather than requiring push-to-talk, the system proactively monitors speech activity to decide when to keep listening and when to treat an utterance as complete: turns are segmented by voice-activity detection (200~ms pre-roll, 500~ms hangover; energy-based in the browser, WebRTC on desktop), so the trainee never issues an explicit turn signal. To keep response onset fast once a turn is judged complete, the server streams tokens and synthesizes each sentence as it finishes, so the persona replies about 1.5~s after the trainee stops speaking, measured across a session (recognition about 0.3~s, first-sentence generation about 1.1~s, then synthesis). The current implementation infers completion from silence alone; handling genuine mid-turn interruptions and backchannels is an open problem we return to in future work.

\vspace{0.5em}
\noindent
\textbf{Persistent Persona State.}
The persona's identity is not fixed. At the loop's center is a
fourteen-dimensional model of the persona's social-emotional condition that
persists across the session rather than resetting each turn. Each turn, a RoBERTa
classifier~\cite{Liu2019RoBERTa} fine-tuned on GoEmotions (\texttt{SamLowe/roberta-base-go\_emotions})~\cite{demszky2020goemotions} reads the trainee's utterance, and its top detected emotions are mapped to signed, confidence-weighted updates across fourteen state dimensions; a small fixed per-turn drift builds trust on a neutral exchange. Unlike agents that fix the persona's affect or re-derive it only from their own last reply, this state accumulates from how the persona is treated over the whole conversation. A derived behavior label (withdrawn, open, anxious, defensive, engaged, or neutral) conditions both what the agent says and how it is embodied; this state is what connects proactive, low-latency turn-taking to a persona that feels responsive rather than merely fast.

\vspace{0.5em}
\noindent
\textbf{Persona-Grounded Dialogue.}
Persona grounding keeps proactive responsiveness from collapsing into generic responsiveness: the agent must respond quickly while still sounding like the same character, situated in this conversation, at this point in its emotional trajectory. Replies come from a large language model (GPT-4o~\cite{OpenAIGPT4o} by default, providers swappable behind one interface). Each persona is a structured profile of demographics, social background, personality, and disclosure rules; the system ships six and authors new ones at runtime. Because a persona fits in context, we compose the whole profile into the system prompt every turn rather than retrieving fragments, alongside the current emotional state, a trust-gated disclosure policy, and recent history, keeping the persona consistent across a long, unscripted exchange.

\vspace{0.5em}
\noindent
\textbf{Embodied Response Timing.}
Seamlessness also depends on whether voice, face, posture, and gesture appear to belong to the same persona state at the same moment. The server maps the state to eight affect scores sent just before the reply audio, so the avatar's disposition is set as speech begins. Facial expressions follow the affect scores through valence-arousal smoothing~\cite{Russell1980Circumplex}, so it evolves rather than snaps; posture and co-speech gestures from the affect scores and reply-text cues; and lip synchronization phonetically from the synthesized audio, with a server-relayed viseme timeline~\cite{Zhou2018VisemeNet} and amplitude-envelope fallback. Gestures and postures are currently discrete and can occasionally mismatch the verbal content; a fidelity limit we address in future work.

\vspace{0.5em}
\noindent
\textbf{Emotion Inferencing and State Dynamics.}
Each trainee utterance is transcribed by ASR into text before emotion classification; the current pipeline conditions on transcript text alone (prosodic and acoustic affect features remain future work). Our RoBERTa classifier extracts the top 3 emotion categories with confidence scores $s_i$ in [0, 1]. We map these categories onto the 14 continuous state dimensions as discussed in the \textit{Persistant Persona State} section using a predefined sensitivity matrix. Updates are signed and confidence-weighted using the formula: $\Delta d = \sum_{i=1}^{3} \delta_{i,d} \cdot s_i$ ,where each of the top 3 predicted categories contributes a signed per-dimension delta $\delta_{i,d}$ from the sensitivity matrix, scaled by the classifier's confidence $s_i$ in that category, allowing predictions with a higher confidence to exert proportionally more influence on the resulting state update. Supportive categories (e.g., caring, approval) push trust and mood upward while damping stress, whereas judgmental categories (e.g., contempt, anger) pull trust down and elevate distress. A subtle per-turn baseline drift toward higher trust and lower stress simulates conversational momentum, before all values are clamped to [0.0, 5.0]. The resulting state vector deterministically maps to one of six coarse behavior tags via priority thresholds on trust, stress, mood, anxiety, anger, and engagement (withdrawn, open, anxious, defensive, engaged, or neutral). This completed state vector is used by injection into the dynamic system prompt to condition LLM dialogue generation, and by normalizing into eight Plutchik scores ([0.0, 1.0]) to drive real-time facial blendshapes and gesture cues before avatar speech begins.

\section{Conclusion and Future Work}

\sysname\ is a voice-driven interaction architecture, built around persona-grounded conversational AI and an embodied virtual agent, that lets users engage in sensitive, high-stakes conversations with the agent through natural, two-way spoken interaction. It moves interaction with AI personas beyond scripted role-play and static simulation, toward repeatable, seamless exchanges with agents that respond in real time and adapt to how they are treated. Alongside continued development and evaluation of \sysname, including a thorough user study of people's experiences with the system, our long-term future work is threefold.

\vspace{0.5em}
\noindent
\textbf{Immersive VR deployment.}
The first direction extends \sysname\ into a VR-based experience. Delivering the agent through a head-mounted display situates the user's full body within the scene rather than behind a screen: the agent could perceive and react to posture, gaze, and other non-verbal cues, and the user's own body language becomes part of the interaction. This would let us study whether immersion changes engagement with an emotionally responsive persona compared with the current screen-based deployment.

\vspace{0.5em}
\noindent
\textbf{More realistic embodied animation.}
The current system animates the agent with affect-driven facial expression, posture and gestures cued from dialogue, and audio-driven lip synchronization. We aim to develop more naturalistic, continuous facial and bodily animation that better mirrors the subtle, moment-to-moment non-verbal behavior of face-to-face interaction, reducing the occasional mismatches between verbal content and expression observed in our case study.

\vspace{0.5em}
\noindent
\textbf{Richer emotion modeling and session scoring.}
The agent already maintains a multi-dimensional emotional state that shifts with how the user communicates, reported as a per-session summary. We plan to deepen this model so the agent's affect more faithfully reflects its evolving disposition toward the user, and to build on that signal a tool that scores a session, surfacing how it navigated rapport, openness, and attunement to support reflective feedback rather than a prescriptive assessment of the user.

\begin{acks}
This work was supported by California HIV/AIDS Research Program (CHRP). We acknowledge the valuable feedback and constructive comments provided by the anonymous reviewers.
\end{acks}

\balance
\bibliographystyle{ACM-Reference-Format}
\bibliography{references}

@inproceedings{demszky2020goemotions,
  title={GoEmotions: A dataset of fine-grained emotions},
  author={Demszky, Dorottya and Movshovitz-Attias, Dana and Ko, Jeongwoo and Cowen, Alan and Nemade, Gaurav and Ravi, Sujith},
  booktitle={Proceedings of the 58th annual meeting of the association for computational linguistics},
  pages={4040--4054},
  year={2020}
}

@article{chen2024oscars,
  title={The oscars of ai theater: A survey on role-playing with language models},
  author={Chen, Nuo and Wang, Yan and Deng, Yang and Li, Jia},
  journal={arXiv preprint arXiv:2407.11484},
  year={2024}
}

@misc{Defossez2024Moshi,
      title={Moshi: a speech-text foundation model for real-time dialogue}, 
      author={Alexandre D\'efossez and Laurent Mazar\'e and Manu Orsini and Am\'elie Royer and Patrick P\'erez and Herv\'e J\'egou and Edouard Grave and Neil Zeghidour},
      year={2024},
      eprint={2410.00037},
      archivePrefix={arXiv},
      primaryClass={eess.AS},
      url={https://arxiv.org/abs/2410.00037}, 
}

@online{OpenAIRealtimeAPI,
    author = {OpenAI},
    title = {Introducing the Realtime API},
    year = {2024},
    url = {https://openai.com/index/introducing-the-realtime-api/},
    note = {Accessed on July 10, 2026}
}

@inproceedings{Liu2025Interruptions,
    author = {Liu, Chao and Su, Mingyang and Xiang, Yan and Huang, Yuru and Yang, Yiqian and Zhang, Kang and Fan, Mingming},
    title = {Toward Enabling Natural Conversation with Older Adults via the Design of LLM-Powered Voice Agents that Support Interruptions and Backchannels},
    year = {2025},
    isbn = {9798400713941},
    publisher = {Association for Computing Machinery},
    address = {New York, NY, USA},
    url = {https://doi.org/10.1145/3706598.3714228},
    doi = {10.1145/3706598.3714228},
    booktitle = {Proceedings of the 2025 CHI Conference on Human Factors in Computing Systems},
    pages = {1--22},
    numpages = {22},
    location = {Yokohama, Japan},
    series = {CHI '25}
}

@inproceedings{Ding2022TalkTive,
    author = {Ding, Zijian and Kang, Jiawen and Ho, Tinky Oi Ting and Wong, Ka Ho and Fung, Helene H. and Meng, Helen and Ma, Xiaojuan},
    title = {TalkTive: A Conversational Agent Using Backchannels to Engage Older Adults in Neurocognitive Disorders Screening},
    year = {2022},
    publisher = {Association for Computing Machinery},
    address = {New York, NY, USA},
    url = {https://doi.org/10.1145/3491102.3502005},
    doi = {10.1145/3491102.3502005},
    booktitle = {Proceedings of the 2022 CHI Conference on Human Factors in Computing Systems},
    pages = {1--19},
    numpages = {19},
    location = {New Orleans, LA, USA},
    series = {CHI '22}
}

@article{Skantze2021,
    title = {Turn-taking in Conversational Systems and Human-Robot Interaction: A Review},
    volume = {67},
    issn = {0885-2308},
    url = {https://doi.org/10.1016/j.csl.2020.101178},
    doi = {10.1016/j.csl.2020.101178},
    pages = {101178},
    journaltitle = {Computer Speech \& Language},
    shortjournal = {Comput. Speech Lang.},
    author = {Skantze, Gabriel},
    date = {2021-05},
}

@inproceedings{Radford2023Whisper,
    title={Robust Speech Recognition via Large-Scale Weak Supervision},
    author={Radford, Alec and Kim, Jong Wook and Xu, Tao and Brockman, Greg and McLeavey, Christine and Sutskever, Ilya},
    booktitle={Proceedings of the 40th International Conference on Machine Learning},
    volume={202},
    pages={28492--28518},
    year={2023},
    organization={PMLR}
}

@online{HexgradKokoro,
    author = {hexgrad},
    title = {Kokoro-82M},
    year = {n.d.},
    url = {https://huggingface.co/hexgrad/Kokoro-82M},
    note = {Accessed on July 10, 2026}
}

@misc{Liu2019RoBERTa,
      title={RoBERTa: A Robustly Optimized BERT Pretraining Approach}, 
      author={Liu, Yinhan and Ott, Myle and Goyal, Naman and Du, Jingfei and Joshi, Mandar and Chen, Danqi and Levy, Omer and Lewis, Mike and Zettlemoyer, Luke and Stoyanov, Veselin},
      year={2019},
      eprint={1907.11692},
      archivePrefix={arXiv},
      primaryClass={cs.CL},
      url={https://arxiv.org/abs/1907.11692}, 
}

@online{OpenAIGPT4o,
    author = {OpenAI},
    title = {Hello GPT-4o},
    year = {2024},
    url = {https://openai.com/index/hello-gpt-4o/},
    note = {Accessed on July 10, 2026}
}

@article{Russell1980Circumplex,
    title = {A Circumplex Model of Affect},
    volume = {39},
    issn = {0022-3514},
    url = {https://doi.org/10.1037/h0077714},
    doi = {10.1037/h0077714},
    pages = {1161--1178},
    number = {6},
    journaltitle = {Journal of Personality and Social Psychology},
    shortjournal = {J. Pers. Soc. Psychol.},
    author = {Russell, James A.},
    date = {1980},
}

@article{Zhou2018VisemeNet,
    title = {VisemeNet: Audio-Driven Animator-Centric Speech Animation},
    volume = {37},
    issn = {0730-0301},
    url = {https://doi.org/10.1145/3197517.3201292},
    doi = {10.1145/3197517.3201292},
    pages = {161},
    number = {4},
    journaltitle = {ACM Transactions on Graphics},
    shortjournal = {ACM Trans. Graph.},
    author = {Zhou, Yang and Xu, Zhan and Landreth, Chris and Kalogerakis, Evangelos and Maji, Subhransu and Singh, Karan},
    date = {2018-08},
}

\end{document}